\documentclass[a4paper,twoside]{article}

\usepackage{epsfig}
\usepackage{subcaption}
\usepackage{calc}
\usepackage{amssymb}
\usepackage{amstext}
\usepackage{amsmath}
\usepackage{amsthm}
\usepackage{multicol}
\usepackage{pslatex}
\usepackage{apalike}
\usepackage{algorithm2e}
\usepackage[bottom]{footmisc}
\usepackage{SCITEPRESS}     

\usepackage{booktabs}
\usepackage{multirow}
\usepackage[table]{xcolor}

\begin{document}

\title{Federated Learning for Malware Image Classification \\under Data Heterogeneity}


\author{\authorname{Victor Taiwo\sup{1}, Cemal Nişan\sup{1}, Muhammad Athallah\sup{2}, M.~Emre Gürsoy\sup{1}, and Öznur Özkasap\sup{1}}
\affiliation{\sup{1}Department of Computer Engineering, Koç University, Istanbul, Türkiye}
\affiliation{\sup{2}Informatics Institute, Istanbul Technical University, Istanbul, Türkiye}
\email{\{vtaiwo25, cnisan21, emregursoy, oozkasap\}@ku.edu.tr, athallah24@itu.edu.tr}
}


\abstract{Malware image classification (MIC), in which binary files are converted into visual representations and deep neural networks are trained to identify malware families, has recently emerged as an effective approach for malware detection. Federated learning (FL) enables malware image classifiers to be trained collaboratively by multiple clients without sharing raw data. However, a fundamental challenge in FL is data heterogeneity (non-IID), which is particularly relevant but understudied in the malware context. In this paper, we address this gap via three contributions. First, we present an empirical study of federated MIC under varying degrees of data heterogeneity across three popular malware image datasets (MalImg, Virus-MNIST, and MalNet). We show that while FL achieves strong accuracy under IID conditions, both accuracy and training stability degrade as data becomes increasingly non-IID. Second, we introduce a set of round-level statistics that characterize the data contributed by participating clients in each FL round, and demonstrate that class entropy (diversity of class labels present in a given round) exhibits strong positive correlation with global model accuracy. Third, leveraging this insight, we propose an entropy-aware client selection strategy that filters candidate client subsets based on class entropy, and show that our strategy outperforms standard random selection. Our results highlight the role of client composition in federated MIC and underscore the importance of distribution-aware classifier training under data heterogeneity.}

\onecolumn \maketitle \normalsize \setcounter{footnote}{0} \vfill

\section{Introduction} \label{sec:Introduction} 

Malware continues to pose a major threat to modern computing systems, with attacks growing in scale, diversity, and sophistication. Deep learning-based approaches are actively being developed for malware detection and classification, particularly using image-based representations of binaries \cite{gibert2020rise,freitas2022malnet,brosolo2024sok}. In this line of work, often termed malware image classification (MIC), binary files are transformed into visual representations (images) and vision models (e.g., convolutional neural networks) are trained to identify structural patterns associated with different malware families \cite{nataraj2011malware,vasan2020imcfn,vasan2020image,kumar2022dtmic,tekerek2022novel}. 

Despite their effectiveness, conventional deep learning approaches require aggregating data from multiple sources into a single location. In the context of malware classification, such data collection raises significant privacy, security, and regulatory concerns, since the raw data of a client may reveal sensitive information about installed software, system vulnerabilities, and malware infections. Here, federated learning (FL) \cite{mcmahan2017communication} provides a compelling solution by enabling clients to collaboratively train a malware classifier without sharing their data. In FL, clients locally train models using their own data and transmit only the updated model weights to the FL server that coordinates the training process, aggregates local updates and computes the weights of the global model. Indeed, successful approaches applying FL to malware image classification have recently been proposed \cite{abdel2022efficient,ambekar2025fasnet}.

However, a fundamental challenge in FL is data heterogeneity, where client data is non-independent and identically distributed (non-IID). This issue is particularly pronounced in malware-related applications, as different client devices (e.g., personal devices running different operating systems, enterprise systems, IoT devices) are exposed to distinct threats and malware, resulting in highly imbalanced data and skewed class distributions. While non-IID data is a known challenge in FL \cite{li2020fedProx,li2022federated,karimireddy2020scaffold,mcmahan2017communication}, there is limited coverage or consideration of it in the MIC context. 

To address this gap, in this paper, we first present an extensive empirical study of federated MIC under varying degrees of data heterogeneity (non-IID) and datasets. Using three popular malware image datasets (MalImg, Virus-MNIST, and MalNet), we show that FL can achieve strong performance under IID conditions, but its accuracy and stability degrade as data becomes increasingly non-IID. Second, we introduce a set of \textit{round-level statistics} (metrics) to characterize the data contributed by participating clients in each FL round, and demonstrate that these statistics (in particular, \textit{class entropy}) are correlated with model accuracy. Third, we leverage this insight to design an entropy-aware client selection strategy that improves the accuracy of federated malware image classifiers over standard random selection. Overall, our methods and results provide new insights into the role of client composition in FL of malware image classifiers, and highlight the importance of distribution-aware training strategies under data heterogeneity.

The rest of this paper is organized as follows. We survey related work in Section \ref{sec:RelatedWork}. Preliminaries for MIC, FL, and our problem statement are provided in Section \ref{sec:Preliminaries}. Our methods, including the proposed round-level statistics and entropy-aware client selection strategy, are described in Section \ref{sec:Methodology}. Experiment results and related analyses are presented in Section \ref{sec:Results}, followed by conclusions in Section \ref{sec:Conclusion}.

\section{Related Work} \label{sec:RelatedWork}

In this section, we review FL approaches for malware detection and classification. We start with a broader scope, including different data modalities (not only image-based methods), and then focus on malware images, which are most directly relevant to our work. 

Several FL methods have been proposed in the literature for malware detection and classification. An FL-based framework for detecting infected Internet of Things (IoT) devices through network traffic, employing the federated averaging (FedAvg) algorithm to aggregate local model updates, was proposed in \cite{Rey2021FederatedLF}. In \cite{Amjath2024GraphRF}, application structure and behavior were modeled using function call graphs, and malware detection was performed using two FL-based graph neural network models: Fed-MalGCN and Fed-MalGAT. In \cite{DAngelo2023PrivacypreservingMD}, a Markov chain and association rule-based classifier was adapted to the FL setting to classify IoT malware. In \cite{nobakht2024sim}, a lightweight one-dimensional CNN with optimized hyperparameters was proposed for IoT malware detection. A concept drift-resilient FL framework named FL-MalDrift was proposed in \cite{patel2025fl}, which integrates lightweight on-device drift detection. Ransomware detection using FL, which is a specific type of malware, was the focus of studies in \cite{vehabovic2023federated} and \cite{lan2025hfl}. Another recent work studied the robustness aspect and highlighted that FL-based malware detectors remain vulnerable to label flipping attacks \cite{eslamnejad2025federated}. Two attack strategies and two density-based defense mechanisms were introduced to counter label flipping attacks.

A few works within the domain of malware detection explored the challenges posed by non-IID data. FedHGCDroid \cite{Jiang2022FedHGCDroidAA} is an FL framework for Android malware classification. It combines HGCDroid, a malware classification model, with FedAdapt, a contribution degree-based adaptive training mechanism that employs attention and meta-learning to improve model resilience to data heterogeneity. In addition, FedDRC \cite{Jiang2024FedDRCAR} was proposed to enhance robustness against spatiotemporal heterogeneity of Android malware. This framework pairs an API encoding mechanism based on functional semantics with an information bottleneck-based local model. 

The works discussed above do not use malware images, but rather the likes of statistical features extracted from network traffic or graph representations of function calls; therefore, their data modality is fundamentally different from ours. Among works focusing on malware images is \cite{abdel2022efficient}, which proposed FED-MAL, an FL framework for MIC in the IoT setting. It utilizes a custom lightweight CNN backbone as the local model and achieves an accuracy of 91.33\% on the Virus-MNIST dataset. Another CNN-based FL approach was proposed in \cite{ullah2023privacy}, in which raw malware binaries were converted into color images, and a data augmentation method was utilized to balance malware data during local training. In \cite{ambekar2025fasnet}, Siamese networks were integrated as feature extractors for MIC in the FL setting. This approach was supplemented with adversarial training to improve model robustness against evasion attacks. Finally, in \cite{Ciaramella2024AnAF}, the authors evaluated the impact of batch normalization on the FedAvg algorithm and provided a comparison with group normalization.

While these works demonstrate the effectiveness of federated MIC, they focus primarily on model architecture, accuracy, and adversarial robustness under homogeneous or mildly heterogeneous data distributions. Critically, they do not perform a systematic evaluation and improvement regarding the impacts of data heterogeneity on classification performance. In this work, we address this gap, studying the impact of non-IID data specifically on federated MIC. As with prior works, we utilize the FedAvg algorithm for aggregation, and similar to \cite{Jiang2022FedHGCDroidAA,Jiang2024FedDRCAR}, we simulate the non-IID setting using a Dirichlet distribution.

\section{Preliminaries and Problem Statement} \label{sec:Preliminaries}

\subsection{System Model and Notation}

Consider an FL setup with $K$ clients, where each client has a local dataset of malware image samples denoted by $D_k = \{(x_{i,k}, y_{i,k})\}_{i=1}^{n_k}$. Here $x_{i,k} \in \mathbb{R}^d $ is an image generated from a program's binary, $y_{i,k} \in  \{1, 2, ..., L\}$ is the ground truth class label of $x_{i,k}$ (malware type or benign), and $n_k$ is the number of samples in $D_k$. We denote by $n$ the total number of samples across all clients, i.e., $n = \sum_{k=1}^Kn_k$. Each image $x_{i,k}$ is a visual representation of a program's binary (malware or benign). There are multiple ways in which the images can be constructed, e.g., the binary is first processed as a vector of 8-bit unsigned integers, then converted to a 2D array, and then visualized as a grayscale image \cite{nataraj2011malware,noever2021virus}. 

Malware image classification (MIC) can be formulated as a supervised multi-class classification problem, where the goal is to learn a model $f_w: \mathbb{R}^d \rightarrow \{1, 2,..., L \}$ that maps image representations of binaries to their corresponding class labels. The model parameters $w$ can be obtained by minimizing:  
\begin{equation} \label{eq:train}
\min_{w} \; \frac{1}{n} \sum_{k=1}^{K} \sum_{i=1}^{n_k} \ell(f_w(x_{i,k}), y_{i,k})
\end{equation}
where $\ell(\cdot)$ denotes the loss function (e.g., cross-entropy loss). 

In a traditional centralized learning approach, all clients send their $D_k$ to a central server, who trains $f_w$ according to Equation \ref{eq:train}. While effective, this centralized approach requires clients to share their sensitive data $D_k$ with a central server, which raises privacy concerns. The sharing of $D_k$ is particularly problematic in the context of MIC since the samples in $D_k$ can enable the server to learn which programs are running and whether there are any malware infections on client $k$'s device. The use of FL in MIC eliminates the need to share $D_k$ with a central server, and therefore, it can help address privacy concerns. 

\subsection{Federated Learning for Malware Image Classification} \label{sec:fed_learning}

Federated learning (FL) enables multiple clients to collaboratively train a global model without exchanging their local datasets \cite{mcmahan2017communication,kairouz2021advances,li2020federated}. FL is a round-by-round, iterative process. We denote by $w^t$ the global model at communication round $t$, and by $w_k^t$ the local model parameters of client $k$ at round $t$. At round $t$, the FL server maintains the current global model $w^t$ and performs the following steps:
\begin{enumerate}
    \item A subset of clients $S^t \subseteq \{1, \ldots, K\}$ is selected to participate in round $t$, e.g., by sampling a fraction $C$ of all clients: $|S^t| = C \cdot K$.
    \item The global model $w^t$ is sent to the participating clients.
    \item Each participating client $k \in S^t$ locally updates the model by performing $E$ epochs of training on its local dataset $D_k$, producing updated parameters $w_k^{t+1}$.
    \item The clients send updated models $w_k^{t+1}$ to the server, and the server aggregates received models to obtain the new global model $w^{t+1}$.
\end{enumerate}
This process is repeated either for a fixed number of $T$ rounds or until model convergence. 

In step 3, the local update performed on client $k$'s device minimizes the local objective:
\begin{equation}
    F_k(w) = \frac{1}{n_k} \sum_{i=1}^{n_k} \ell(f_w(x_{i,k}), y_{i,k})
\end{equation}
by performing $E$ epochs of stochastic gradient descent (SGD) initialized at $w^t$. 

For the aggregation step (step 4), we employ Federated Averaging (FedAvg) \cite{mcmahan2017communication}, the de facto aggregation algorithm in FL. In FedAvg, the global model is computed as the weighted average of local models:
\begin{equation}
    w^{t+1} = \sum_{k \in S^t} \frac{n_k}{\sum\limits_{j \in S^t} n_j} \, w_k^{t+1}
\end{equation}
where the contribution of each client is proportional to the size of its local dataset.

\subsection{Problem Statement} \label{sec:ProblemStatement}

A fundamental challenge in FL is data heterogeneity, where client data is not identically and independently distributed (non-IID), i.e., $D_k$ of different clients $k$ differ significantly in terms of cardinality, class distribution, and/or sample characteristics. Prior work has shown that such heterogeneity can significantly degrade model performance, and slow or prevent convergence in algorithms such as FedAvg, FedProx, and SCAFFOLD \cite{mcmahan2017communication,li2020federated,li2022federated,jimenez2025thorough}. 


Non-IID distributions are particularly relevant in federated learning of malware image classifiers. In realistic environments, clients may span personal electronic devices (laptops, smartphones, tablets), enterprise systems, and IoT devices, each exposed to distinct threat landscapes. For example, enterprise systems may encounter targeted or advanced malware families, whereas personal devices are more likely to contain commodity malware such as adware or spyware. Furthermore, different devices run different operating systems and applications; therefore, they are likely to contain different vulnerabilities and be affected by different malware. Consequently, the data distributions across different clients are likely to be imbalanced and non-IID. Data heterogeneity is further exacerbated by the client sampling process in FL: in each round $t$, only a subset of clients $S^t$ participates in training, meaning that the effective training data distribution varies across rounds depending on the participating clients. As a result, the aggregated model update may be biased toward the class distributions present in $S^t$, introducing additional variability and instability in the training process.

Despite the inherently non-IID nature of data in malware-related applications, its impact on federated MIC remains underexplored. In this paper, we aim to fill this gap. More specifically, we formulate and answer the following research questions (RQs):

\textbf{RQ1:} Can FL be used to train accurate malware image classifiers under IID and non-IID data distributions? How does increasing data heterogeneity (non-IID'ness) affect the robustness and accuracy of these models?

\textbf{RQ2:} Can certain round-level characteristics of participating clients, such as aggregate statistical properties of their data, influence or correlate with global model accuracy?

\textbf{RQ3:} Can these round-level statistics be leveraged to design more effective client selection strategies that improve the accuracy of federated malware image classifiers, compared to random client participation?

\section{Methodology} \label{sec:Methodology}

\subsection{System Architecture} \label{sec:Architecture}

\begin{figure}[!b]
    \centering
    \includegraphics[width=.95\columnwidth]{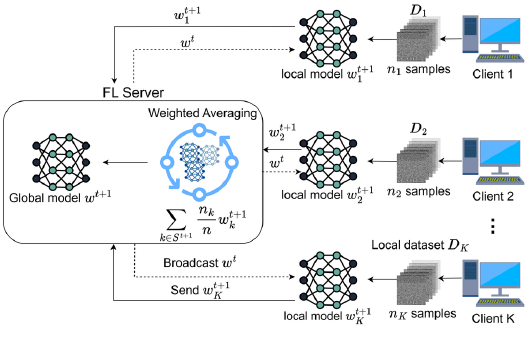}
    \caption{Our system architecture for federated learning of malware image classifiers (MICs).}
    \label{fig:FL}
\end{figure}

We utilize an FL setup consisting of an FL server (aggregator) and $K$ = 50 clients, as illustrated in Figure \ref{fig:FL}. The FL server is responsible for global model initialization, broadcasting the current model to clients, and aggregating the updated models received from clients. Clients are responsible for training local models using their private dataset $D_k$ of malware images, and communicating the updated model weights to the server. We use MobileNetV2 as the model architecture \cite{sandler2018mobilenetv2}. MobileNetV2 is a lightweight convolutional neural network architecture designed for efficient use on resource-constrained devices; therefore, it is well-suited for FL and image data.

\subsection{Malware Image Datasets} \label{sec:Datasets}

\begin{figure}[!t]
    \centering
    \begin{subfigure}{0.45\columnwidth}
    \centering
    \includegraphics[width=0.9\linewidth]{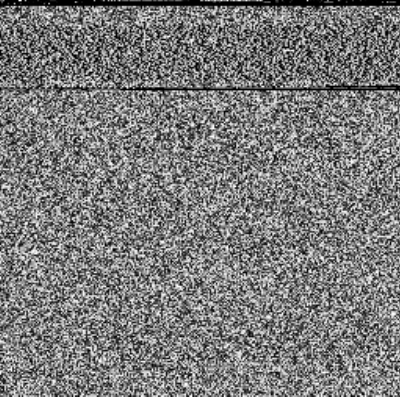}
    \subcaption{}
    \label{fig:allaple}
    \end{subfigure}
    \hspace{4pt}
    \begin{subfigure}{0.45\columnwidth}
    \centering
    \includegraphics[width=0.9\linewidth]{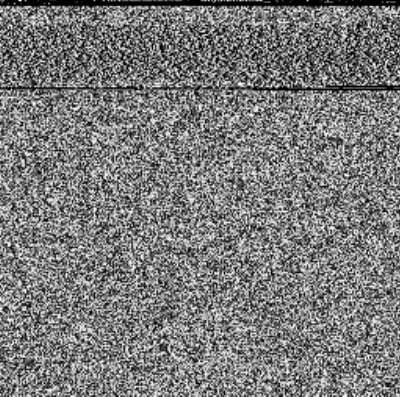}
    \subcaption{}
    \label{fig:allaple_b}
    \end{subfigure}
\\ \vspace{6pt}
    \begin{subfigure}{0.45\columnwidth}
    \centering
    \includegraphics[width=0.9\linewidth]{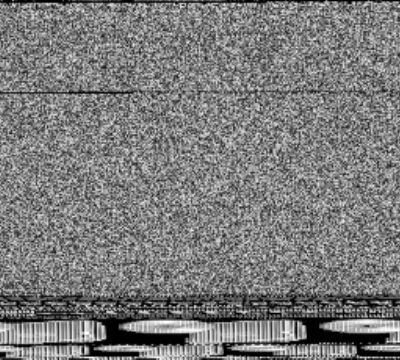}
    \subcaption{}
    \label{fig:fakerean}
    \end{subfigure}
    \hspace{4pt}
    \begin{subfigure}{0.45\columnwidth}
    \centering
    \includegraphics[width=0.8\linewidth]{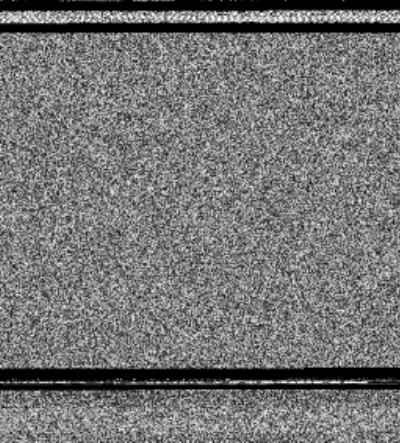}
    \subcaption{}
    \label{fig:obfuscator}
    \end{subfigure}
    \caption{Sample images from the MalImg dataset. (a) and (b) are samples from the \textit{Allaple.A} family of network worms. (c) is a sample from the \textit{Fakerean} trojan, and (d) is a sample from the \textit{Obfuscator.AD!} trojan.}
    \label{fig:malware}
\end{figure}

We used three publicly available malware image datasets in our work: MalImg, Virus-MNIST, and MalNet. Sample images from the MalImg dataset are provided in Figure \ref{fig:malware}. It can be observed that samples from the same malware family \textit{Allaple.A} (Figure \ref{fig:malware}a and \ref{fig:malware}b) are visually similar, whereas samples from the other families (Figure \ref{fig:malware}c and \ref{fig:malware}d) are visually distinct.

MalImg \cite{nataraj2011malware} is a widely used malware image dataset consisting of grayscale images generated from malware binaries. The dataset contains a total of 9,339 samples spanning 25 malware families, such as Allaple, Yuner, Instantaccess, and Fakerean. We used an 80\%-20\% train-test split to ensure sufficient representatives from each family in our test set. 

Virus-MNIST \cite{noever2021virus} is a dataset of 51,880 grayscale images across 10 classes generated from Windows Portable Executable files. It consists of 2,516 samples from benign programs and 49,364 malware samples divided into 9 malware families. We maintained the 85\%-15\% train-test split implemented by the authors.

MalNet \cite{freitas2022malnet} is a large-scale malware image dataset derived from Android applications (APK files). It contains hundreds of thousands of samples spanning numerous malware families, where each application is converted into an image representation based on its bytecode. In addition to grayscale representations, it incorporates an additional semantic coloring of each pixel based on its function. We performed our experiments on a subset of this dataset consisting of 45,385 samples from the ten largest classes. Furthermore, visual inspection revealed that informative patterns were primarily along the vertical dimension, so we cropped the images to 32×256 to improve model convergence speed, allowing us to better focus on the impact of IID versus non-IID. We used a 70\%-30\% train-test split. 

\subsection{Non-IID Simulation and Hyperparameters} \label{sec:non-IID}

\begin{figure*}[!t]
    \centering
    \includegraphics[width=.98\textwidth]{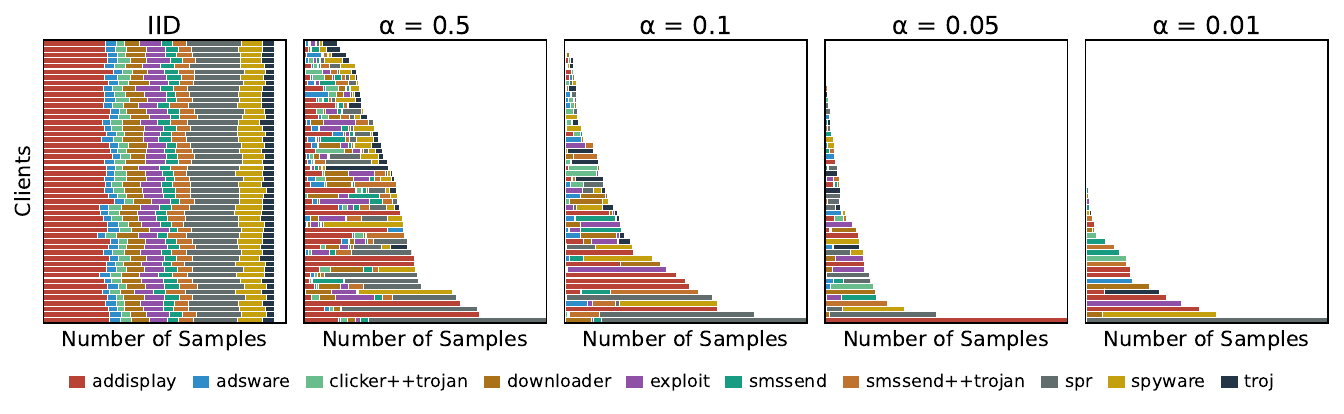}
    \vspace{-2pt}
    \caption{Distribution of MalNet dataset among 50 clients under varying degrees of non-IID, controlled by the $\alpha$ parameter. Each class is represented by a different color. It can be observed that $\alpha$ = 0.5 is closer to IID, as clients have more similar numbers of samples, and their samples span multiple different classes. In contrast, when $\alpha$ = 0.01, clients have very different dataset sizes and their samples are predominantly from one or two classes.}
    \label{fig:non-IID}
\end{figure*}

To simulate non-IID data distributions across clients, we adopt the commonly used Dirichlet partitioning strategy. Specifically, for each class $l \in \{1, \ldots, L\}$, we sample a $K$-dimensional proportion vector $P_l = (p_{l,1}, \ldots, p_{l,K}) \sim \text{Dir}(\alpha)$, where $\alpha > 0$ is the concentration parameter controlling the degree of non-IID'ness. Each $p_{l,k}$ determines the fraction of samples of class $l$ assigned to client $k$. Smaller values of $\alpha$ lead to highly skewed distributions where clients predominantly observe a small subset of classes, while larger values of $\alpha$ yield more balanced distributions closer to IID. Figure \ref{fig:non-IID} provides visualizations of the data distributions for different values of $\alpha$ on the MalNet dataset.

We used $E$ = 3 local epochs, fraction $C = 0.1$, and $T$ = 100 communication rounds for training. Local models were trained with the SGD optimizer, using learning rate $\eta = 10^{-2}$, and momentum $\mu = 0.9$. We did not utilize weight decay as baseline experiments without it already showed satisfactory performance. A batch size of 10 for MalImg and 32 for MalNet and Virus-MNIST is used, based on our empirical observations regarding model convergence.

\subsection{Round-Level Statistics} \label{sec:Statistics}

Only a subset of clients $S^t \subseteq \{1, \ldots, K\}$ participate in each round $t$ of FL. As discussed in Section \ref{sec:ProblemStatement}, this induces variability in the effective training data amount and distribution across rounds, especially in non-IID scenarios. To analyze the effects of this variability and which aspects of the variability affect model accuracy the most, we propose a set of \textit{round-level statistics} that capture different properties of the data contributed by clients $S^t$. Our statistics are designed to consider complementary aspects of the data distribution at each round, enabling us to analyze how client composition influences model accuracy under non-IID data.

\subsubsection{Total Samples}

We define the total number of samples $N^t$ available in round $t$ as:
\begin{equation}
    N^t = \sum_{k \in S^t} n_k
\end{equation}
This metric captures the overall amount of data used to compute the global update in round $t$. In general, larger values of $N^t$ are expected to yield more stable and accurate models, since the model learns from a larger number of samples. However, under non-IID settings, large $N^t$ may still correspond to highly imbalanced data if the majority of the samples come from a few classes.

\subsubsection{Median Samples}

We compute the median of the set $\{n_k : k \in S^t\}$, which reflects the median dataset size among participating clients. Unlike $N^t$, this metric is less sensitive to outliers, i.e., few clients with very large or very small $n_k$. 

\subsubsection{Sample Entropy}

To quantify how evenly the data is distributed across participating clients, we compute the Shannon entropy of the sample distribution. Let:
\begin{equation}
p_k^t = \frac{n_k}{N^t}, \quad k \in S^t
\end{equation}
denote the fraction of samples contributed by client $k$ in round $t$. The sample entropy is then defined as:
\begin{equation}
    H_{\text{sample}}^t = - \sum_{k \in S^t} p_k^t \log_2 p_k^t
\end{equation}
Higher values of $H_{\text{sample}}^t$ indicate a more balanced contribution across clients, while lower values indicate that a small number of clients dominate the update. Such dominance can bias the aggregated model toward specific client distributions, which can be problematic under non-IID data.

\subsubsection{Class Entropy}

We note that sample entropy considers only the \textit{amount} of data, not the class the data belongs to. To measure the diversity of class labels in the data available in round $t$, we compute the Shannon entropy of the class distribution. Let $n^{(l,t)}$ denote the number of samples of class $l$ in round $t$, and define:
\begin{equation}
q_l^t = \frac{n^{(l,t)}}{N^t}, \quad l \in \{1, \ldots, L\}
\end{equation}
The class entropy is given by:
\begin{equation}
    H_{\text{class}}^t = - \sum_{l=1}^{L} q_l^t \log_2 q_l^t
\end{equation}
The key difference between class entropy and sample entropy is that class entropy captures the diversity of class labels present in the participating clients. Low class entropy indicates that the round is dominated by a small number of classes, which can lead to biased model updates and reduced generalization performance, as the model will not see sufficiently many samples from other classes.

\subsubsection{Average Class Entropy}

We also consider the average class entropy across individual clients. For each client $k \in S^t$, let $n_k^{(l,t)}$ denote the number of samples of class $l$ in client $k$'s dataset, and define:
\begin{equation}
q_{k,l}^t = \frac{n_k^{(l,t)}}{n_k}    
\end{equation}
Then, the class entropy of client $k$ at round $t$ is:
\begin{equation}
    H_{\text{class}}^{k,t} = - \sum_{l=1}^{L} q_{k,l}^t \log_2 q_{k,l}^t
\end{equation}
Finally, the average class entropy in round $t$ is:
\begin{equation}
    \bar{H}_{\text{class}}^t = \frac{1}{|S^t|} \sum_{k \in S^t} H_{\text{class}}^{k,t}
\end{equation}
This metric reflects how diverse the class distributions are within individual clients. Low values indicate that clients' local datasets $D_k$ are concentrated on a few classes, which can amplify heterogeneity effects during FL.

\subsubsection{Weighted Average Class Entropy}

We define a weighted version of average class entropy that accounts for client dataset sizes:
\begin{equation}
    \widehat{H}_{\text{class}}^t = \frac{1}{N^t} \sum_{k \in S^t} n_k \, H_{\text{class}}^{k,t}
\end{equation}
This metric captures the overall class diversity while giving higher weight to clients with larger datasets, since their updates are likely to be more dominant in FL. It serves to balance the class distribution and sample size.

\subsection{Entropy-Aware Client Selection} \label{sec:EntropyAwareSelection}

We empirically observed that class entropy $H_{\text{class}}^t$, in particular, exhibits a strong and consistent correlation with model accuracy. To answer RQ3 by turning this observation into an actionable insight, we propose an entropy-aware client selection strategy. The key intuition of our strategy is that rounds with higher class entropy tend to produce more representative global updates; therefore, explicitly favoring such client subsets in FL will improve model accuracy.

Our proposed strategy operates as follows. At each round $t$:
\begin{enumerate}
    \item The FL server generates a large pool of candidate client subsets, each of size $C \cdot K$, by sampling combinations of clients uniformly at random among all $K$ clients.
    \item For each candidate subset, the statistics defined in Section \ref{sec:Statistics} are computed. 
    \item The candidate subsets are filtered by discarding: (i) those subsets with total number of samples $N^t$ below the 25th percentile, and (ii) those with class entropy below the 95th percentile. 
    \item Among the remaining candidates, one of them is randomly selected to become the $S^t$ that will participate in round $t$. 
\end{enumerate}
The filtering in the third step ensures that: (i) there will be sufficient data per round of training, and (ii) following the intuition from the previous section, high class entropy (among the top 5\%) will be enforced among the possible candidates. The selection in step 4 is non-deterministic and without replacement, so that our strategy does not always select the same clients every round.

\section{Results and Analysis} \label{sec:Results}

In this section, we experimentally analyze and discuss the answers to the three RQs stated in Section \ref{sec:ProblemStatement}. We dedicate Section \ref{sec:RQ1} to RQ1, Section \ref{sec:RQ2} to RQ2, and Section \ref{sec:RQ3} to RQ3. Lastly, we perform more detailed hyperparameter analyses in Section \ref{sec:Hyperparameters}.

\subsection{Impact of Non-IID on Model Accuracy and Convergence} \label{sec:RQ1}

\begin{table*}[!t]
\centering
\caption{Test accuracy (\%) under varying degrees of non-iid}
\label{tab:results}
\begin{tabular}{lccccc}
\toprule
Dataset & IID & $\alpha=0.5$ & $\alpha=0.1$ & $\alpha=0.05$ & $\alpha=0.01$ \\
\midrule
MalImg     & $99.04 {\scriptstyle\,\pm\, 0.28}$ & $96.83 {\scriptstyle\,\pm\, 4.00}$ & $83.92 {\scriptstyle\,\pm\, 9.03}$ & $74.04 {\scriptstyle\,\pm\, 12.45}$ & $47.63 {\scriptstyle\,\pm\, 22.96}$ \\
Virus-MNIST & $91.97 {\scriptstyle\,\pm\, 0.42}$ & $86.35 {\scriptstyle\,\pm\, 2.19}$ & $65.69 {\scriptstyle\,\pm\, 11.74}$ & $59.24 {\scriptstyle\,\pm\, 13.44}$ & $22.99 {\scriptstyle\,\pm\, 0.12}$ \\
MalNet     & $84.30 {\scriptstyle\,\pm\, 0.17}$ & $78.53 {\scriptstyle\,\pm\, 1.78}$ & $55.77 {\scriptstyle\,\pm\, 13.32}$ & $50.35 {\scriptstyle\,\pm\, 5.17}$ & $14.31 {\scriptstyle\,\pm\, 11.74}$ \\
\bottomrule
\end{tabular}
\end{table*}

To answer RQ1, we evaluate the performance of our FL setup under varying degrees of non-IID, simulated using Dirichlet distributions with concentration parameter $\alpha \in \{0.5, 0.1, 0.05, 0.01\}$. We also include results under IID data as a point of comparison. (The data distributions are exemplified in Figure \ref{fig:non-IID}.) Table \ref{tab:results} summarizes the accuracy values of the global model in each setting, across the MalImg, Virus-MNIST, and MalNet datasets.

Results in Table \ref{tab:results} show a consistent degradation in model accuracy as the degree of non-IID'ness increases. Across all datasets, models achieve the highest accuracy under IID conditions, and their performance decreases as $\alpha$ becomes smaller. When $\alpha = 0.5$, which represents a moderate level of heterogeneity, the reduction in accuracy remains relatively limited (approximately $4.5\%$ on average), indicating that FedAvg can tolerate mild heterogeneity. However, as $\alpha$ decreases further, the impact becomes more pronounced. In the extreme case of $\alpha = 0.01$, the model accuracy drops to less than half of the IID baseline. 

We would like to highlight that under IID conditions, highly accurate malware image classifiers can be trained using FL. For example, as provided in Table \ref{tab:results}, we achieve accuracies exceeding $99\%$ on MalImg and $91\%$ on Virus-MNIST. These results are comparable to those reported in the original dataset papers and to those obtained under centralized training. Therefore, the primary challenge does not stem from FL itself, but rather the presence of non-IID data distributions.

\begin{figure*}[!t]
    \centering
    \includegraphics[width=.95\textwidth]{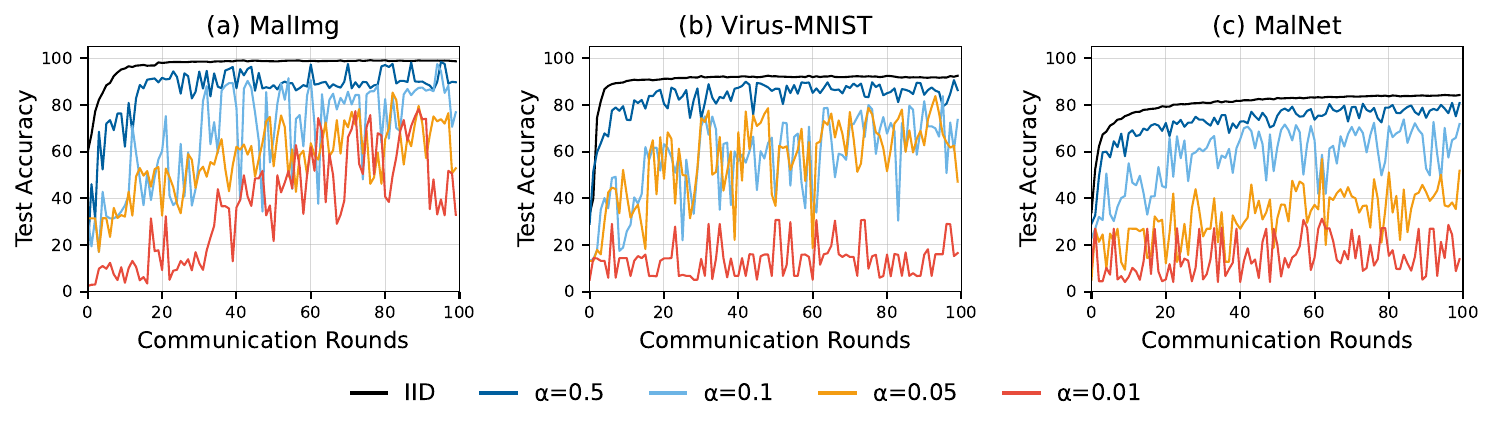}
    \vspace{-2pt}
    \caption{Accuracy versus rounds on different datasets under varying $\alpha$}
    \label{fig:non-iid accuracy}
\end{figure*}

In addition to reduced accuracy, we observe increased instability in the training process as heterogeneity increases. Specifically, the model exhibits greater variability in test accuracy across communication rounds when $\alpha$ is small. This behavior is illustrated in Figure \ref{fig:non-iid accuracy}, where learning curves become increasingly volatile as $\alpha$ is reduced. Under IID conditions, the training process is smooth and stable, with accuracy steadily increasing and quickly converging to a high level across all datasets. For moderate non-IID (e.g., $\alpha = 0.5$), the model still converges to relatively high accuracy (although the final accuracy is slightly lower than that of IID), but with noticeable fluctuations during training. However, as $\alpha$ decreases further, the instability becomes significantly more pronounced. For $\alpha$ = 0.1 and $\alpha$ = 0.05, the accuracy curves exhibit frequent and large oscillations, indicating that successive global updates may move the model in inconsistent directions. The final model accuracy at the end of training is also noticeably lower than IID and $\alpha$ = 0.5. In the extreme case of $\alpha = 0.01$, the training process becomes highly erratic, with accuracy fluctuating substantially and often failing to converge to a stable or accurate model. While some learning occurs on MalImg, the model does not perform better than a random guess on Virus-MNIST or MalNet at any point during training. 

Overall, these behaviors can be attributed to the increasing mismatch between local data distributions and the global objective as $\alpha$ becomes smaller. When client data is highly skewed, local updates computed during each round are strongly biased toward a small subset of classes. As different subsets of clients participate in different rounds, the aggregated updates vary significantly, leading to accuracy oscillations and convergence instability. 

\subsection{Round-Level Statistics and Model Accuracy} \label{sec:RQ2}

Motivated by the reduced accuracy and increased oscillations observed under non-IID data, we investigate how the composition of participating clients in each round influences model performance (RQ2). More specifically, for each round $t$, we compute the round-level statistics of participating $S^t$ using the metrics proposed in Section \ref{sec:Statistics}, and also record the accuracy of the model in that round. Then, we compute the Spearman correlation between each metric and model accuracy, across 100 rounds and 5 independent runs (a total of 500 recordings). Table \ref{tab:correlation} summarizes the results across different datasets and $\alpha$ values. In each column, the metric that achieves the strongest correlation is indicated in bold. Cells shaded with gray background color indicate strong statistical significance: p-value $< 0.05$ in at least 4 of 5 runs. 

\begin{table*}[!t]
\centering
{\fontsize{7pt}{7pt}\selectfont
\renewcommand{\arraystretch}{1.3}
\caption{Correlation of round-level statistics with model accuracy across different datasets and $\alpha$ values}
\label{tab:correlation}
\begin{tabular}{l@{\hspace{0.5cm}}cccc@{\hspace{0.5cm}}cccc@{\hspace{0.5cm}}cccc}
\toprule
& \multicolumn{4}{c}{\textbf{MalImg}} 
& \multicolumn{4}{c}{\textbf{Virus-MNIST}} 
& \multicolumn{4}{c}{\textbf{MalNet}} \\
\cmidrule(lr){2-5} \cmidrule(lr){6-9} \cmidrule(lr){10-13}
\textbf{Metric} & 0.5 & 0.1 & 0.05 & 0.01 
       & 0.5 & 0.1 & 0.05 & 0.01 
       & 0.5 & 0.1 & 0.05 & 0.01 \\
\midrule
Total Samples    & \cellcolor{gray!25}-0.358 & \cellcolor{gray!25}-0.347 & -0.154 & 0.062 & \cellcolor{gray!25}-0.271 & 0.000 & 0.119 & -0.124 & -0.014 & -0.091 & -0.014 & \cellcolor{gray!25}\textbf{0.399} \\
Median Samples   & -0.292 & -0.132 & 0.009 & 0.027 & -0.162 & 0.172 & 0.260 & 0.043 & -0.094 & 0.165 & 0.205 & \cellcolor{gray!25} 0.245 \\
Sample Entropy      & 0.193 & \cellcolor{gray!25}0.356 & 0.291 & -0.001 & \cellcolor{gray!25}0.256 & \cellcolor{gray!25}\textbf{0.311} & \cellcolor{gray!25} 0.350 & \cellcolor{gray!25} 0.307 & -0.023 & \cellcolor{gray!25} 0.400 & \cellcolor{gray!25} 0.423 & 0.214 \\
Class Entropy      & \cellcolor{gray!25}\textbf{0.578} & \cellcolor{gray!25} 0.413 & \textbf{0.375} & \textbf{0.065} & \cellcolor{gray!25}\textbf{0.507} & \cellcolor{gray!25} 0.295 & \cellcolor{gray!25}\textbf{0.459} & \cellcolor{gray!25}\textbf{0.313} & \cellcolor{gray!25}\textbf{0.289} & \cellcolor{gray!25}\textbf{0.589} & \cellcolor{gray!25}\textbf{0.499} & 0.239 \\
Avg. Class Entropy      & \cellcolor{gray!25}0.532 & \cellcolor{gray!25}0.296 & 0.230 & -0.022 & \cellcolor{gray!25}0.430 & 0.045 & 0.061 & 0.094 & \cellcolor{gray!25}0.201 & \cellcolor{gray!25}0.242 & 0.001 & -0.075 \\
Wght. Avg. Class Entropy      & \cellcolor{gray!25} 0.562 & \cellcolor{gray!25}\textbf{0.425} & \cellcolor{gray!25} 0.359 & -0.135 & \cellcolor{gray!25} 0.491 & 0.168 & 0.252 & 0.172 & \cellcolor{gray!25} 0.233 & \cellcolor{gray!25}0.377 & 0.179 & -0.019 \\
\bottomrule
\end{tabular}
}

\end{table*}

Across all datasets and most values of $\alpha$, class entropy $H_{\text{class}}^t$ exhibits the strongest and most consistent positive correlation with model accuracy. In particular, for moderate heterogeneity levels ($\alpha$ = 0.5, 0.1, 0.05), the correlation is both strong and statistically significant across all datasets. This indicates that rounds with more diverse class distributions tend to yield better-performing global models. However, under extreme heterogeneity ($\alpha = 0.01$), this relationship weakens, suggesting that when client data becomes highly skewed, even high-entropy rounds may become insufficient. Sample entropy $H_{\text{sample}}^t$ also shows a generally positive correlation with model accuracy, particularly on Virus-MNIST and MalNet. This suggests that a more balanced contribution of data across clients serve to stabilize model performance and mitigate accuracy degradation. However, its correlation is weaker and less significant than that of class entropy. Similarly, while average class entropy $\bar{H}_{\text{class}}^t$ and its weighted version $\widehat{H}_{\text{class}}^t$ exhibit moderate positive correlations in some cases, they do not outperform class entropy in general. This suggests that the aggregate class distribution across all participating clients (captured by $H_{\text{class}}^t$) is more informative than client-level $\bar{H}_{\text{class}}^t$ and $\widehat{H}_{\text{class}}^t$.

We also study the correlations of Total Samples and Median Samples with model accuracy. Contrary to the common expectation that more data leads to better performance, we often observe a negative correlation between the total number of samples $N^t$ and model accuracy in several configurations, particularly on the MalImg dataset. Furthermore, median number of samples also does not demonstrate a significant correlation with accuracy. These do not imply that larger datasets are inherently detrimental; rather, it reflects the interaction between sample size and data heterogeneity. Due to the inherent skew as shown in Figure~\ref{fig:non-IID}, rounds with large $N^t$ are often dominated by a small number of clients with large but skewed datasets, which reduces both class and sample entropy and leads to worse model updates. Indeed, having sufficiently many samples in a round is often beneficial to improve accuracy, as evidenced by positive correlations especially when $\alpha$ = 0.01 and 0.05.  

\subsection{Evaluation of Entropy-Aware Client Selection} \label{sec:RQ3}

Recall that we proposed an entropy-aware client selection strategy in Section \ref{sec:EntropyAwareSelection}. In this section, we evaluate the effectiveness of our strategy by comparing it against standard random client selection. Hyperparameters are kept the same as in previous experiments, each experiment is repeated five times to ensure statistical reliability, and the average results are reported in Table \ref{tab:client_selection}.

\begin{table}[!t]
\caption{Test accuracy (\%) under random client selection versus our entropy-aware client selection strategy}
\centering
\label{tab:client_selection}
\begin{tabular}{llcc}
\toprule
\textbf{Dataset} & \textbf{$\alpha$} & \textbf{Random} & \textbf{Entr-Aware} \\
\midrule
\multirow{4}{*}{MalImg}
 & 0.5  & $96.83 {\scriptstyle\,\pm\, 4.00}$  & $\mathbf{98.90} {\scriptstyle\,\pm\, 0.09}$ \\
 & 0.1  & $83.92 {\scriptstyle\,\pm\, 9.03}$  & $\mathbf{96.51} {\scriptstyle\,\pm\, 3.64}$ \\
 & 0.05 & $74.04 {\scriptstyle\,\pm\, 12.45}$ & $\mathbf{92.78} {\scriptstyle\,\pm\, 2.95}$ \\
 & 0.01 & $47.63 {\scriptstyle\,\pm\, 22.96}$ & $\mathbf{58.51} {\scriptstyle\,\pm\, 12.43}$ \\
\midrule
\multirow{4}{*}{\small{Virus-MNIST}}
 & 0.5  & $\mathbf{86.35} {\scriptstyle\,\pm\, 2.19}$  & $85.47 {\scriptstyle\,\pm\, 2.97}$ \\
 & 0.1  & $65.69 {\scriptstyle\,\pm\, 11.74}$ & $\mathbf{71.08} {\scriptstyle\,\pm\, 9.90}$ \\
 & 0.05 & $59.24 {\scriptstyle\,\pm\, 13.44}$ & $\mathbf{62.51} {\scriptstyle\,\pm\, 12.27}$ \\
 & 0.01 & $22.99 {\scriptstyle\,\pm\, 0.12}$  & $\mathbf{26.52} {\scriptstyle\,\pm\, 11.28}$ \\
\midrule
\multirow{4}{*}{MalNet}
 & 0.5  & $78.53 {\scriptstyle\,\pm\, 1.78}$  & $\mathbf{78.82} {\scriptstyle\,\pm\, 1.52}$ \\
 & 0.1  & $55.77 {\scriptstyle\,\pm\, 13.32}$ & $\mathbf{67.17} {\scriptstyle\,\pm\, 3.87}$ \\
 & 0.05 & $50.35 {\scriptstyle\,\pm\, 5.17}$  & $\mathbf{54.16} {\scriptstyle\,\pm\, 4.13}$ \\
 & 0.01 & $14.31 {\scriptstyle\,\pm\, 11.74}$ & $\mathbf{24.61} {\scriptstyle\,\pm\, 11.30}$ \\
\bottomrule
\end{tabular}
\end{table}

Across almost all datasets and non-IID levels $\alpha$, the proposed entropy-aware strategy consistently outperforms random selection. The only exception is the Virus-MNIST dataset at $\alpha = 0.5$, where both methods achieve comparable performance (the difference between them is lower than 1\%). The Wilcoxon signed-rank test with a significance level of 0.01 also showed that these results are statistically significant. These results demonstrate that prioritizing class entropy in client selection leads to more effective training and improved model accuracy. Overall, these experiments provide empirical evidence that round-level statistics, particularly class entropy, can be used not only to explain performance variability but also to actively improve the learning of malware image classifiers under non-IID data.

\subsection{Hyperparameter Analyses} \label{sec:Hyperparameters}

We study the effects of the following three parameters on model accuracy: the participating fraction $C$, the number of local training epochs $E$, and the total number of FL clients $K$. We maintain the same hyperparameters used in previous experiments, only varying the specific parameter being tested. Experiments are performed on the MalImg dataset, in IID and non-IID settings with $\alpha$ = 0.1 and 0.01. 

\begin{table}[!t]
\centering
\caption{Impact of participating client fraction ($C$) on model accuracy}
\label{tab:client_participation}
\begin{tabular}{cccc}
\toprule
 \textbf{$C$} & \textbf{IID} & \textbf{$\alpha = 0.1$} & \textbf{$\alpha = 0.01$}  \\
\midrule
0.1 & $99.04 {\scriptstyle\,\pm\, 0.28}$ & $83.92 {\scriptstyle\,\pm\, 9.03}$ & $47.63 {\scriptstyle\,\pm\, 22.96}$ \\
0.2 & $\mathbf{99.20} {\scriptstyle\,\pm\, 0.19}$ & $92.21 {\scriptstyle\,\pm\, 4.40}$ & $51.04 {\scriptstyle\,\pm\, 15.77}$ \\
0.4 & $99.10 {\scriptstyle\,\pm\, 0.06}$ & $91.65 {\scriptstyle\,\pm\, 4.16}$ & $\mathbf{56.75} {\scriptstyle\,\pm\, 19.80}$ \\
0.8 & $98.90 {\scriptstyle\,\pm\, 0.11}$ & $\mathbf{93.65} {\scriptstyle\,\pm\, 4.55}$ & $47.88 {\scriptstyle\,\pm\, 10.76}$ \\
1.0 & $98.94 {\scriptstyle\,\pm\, 0.13}$ & $92.12 {\scriptstyle\,\pm\, 4.22}$ & $41.26 {\scriptstyle\,\pm\, 9.74}$ \\
\bottomrule
\end{tabular}
\end{table}

\textbf{Impact of $C$.} Table \ref{tab:client_participation} summarizes the results under IID and non-IID settings. Under IID conditions, varying $C$ has only a marginal effect on model accuracy, with all configurations achieving consistently high performance (around $99\%$). This indicates that when data is uniformly distributed, the number of participating clients per round is not a critical factor. 

In contrast, we observe more significant variation in model performance in the non-IID settings. When $\alpha$ = 0.1, increasing $C$ initially improves accuracy, with accuracy peaking at $C$ = 0.8, suggesting that involving more clients helps incorporate more diverse data and mitigates heterogeneity. However, further increasing $C$ to $1.0$ does not yield additional gains and slightly degrades performance. A similar but more pronounced trend is observed when $\alpha = 0.01$. Increasing $C$ from $0.1$ to $0.4$ improves accuracy, but beyond this point, performance deteriorates, with $C = 1.0$ resulting in the lowest accuracy. Overall, these results indicate that using the majority or all of the clients does not guarantee better accuracy -- in fact, increasing $C$ can yield diminishing or even negative returns. It should also be noted that an increased $C$ also requires higher communication and computational overhead for the clients, which is undesirable.  

\begin{table}[!t]
\centering
\caption{Impact of the number of local epochs ($E$) on model accuracy}
\label{tab:local_epochs}
\begin{tabular}{cccc}
\toprule
 \textbf{$E$} & \textbf{IID} & \textbf{$\alpha = 0.1$} & \textbf{$\alpha = 0.01$}  \\
\midrule
1  & $98.82 {\scriptstyle\,\pm\, 0.16}$ & $81.56 {\scriptstyle\,\pm\, 12.97}$ & $39.99 {\scriptstyle\,\pm\, 10.36}$ \\
3  & $99.04 {\scriptstyle\,\pm\, 0.28}$ & $83.92 {\scriptstyle\,\pm\, 9.03}$  & $47.63 {\scriptstyle\,\pm\, 22.96}$ \\
5  & $\mathbf{99.22} {\scriptstyle\,\pm\, 0.11}$ & $88.93 {\scriptstyle\,\pm\, 6.80}$  & $\mathbf{51.57} {\scriptstyle\,\pm\, 16.62}$ \\
7  & $98.92 {\scriptstyle\,\pm\, 0.09}$ & $\mathbf{90.05} {\scriptstyle\,\pm\, 5.10}$  & $32.83 {\scriptstyle\,\pm\, 14.30}$ \\
9  & $98.94 {\scriptstyle\,\pm\, 0.16}$ & $87.28 {\scriptstyle\,\pm\, 9.76}$  & $37.12 {\scriptstyle\,\pm\, 5.27}$ \\
11 & $98.86 {\scriptstyle\,\pm\, 0.27}$ & $85.53 {\scriptstyle\,\pm\, 11.97}$ & $31.14 {\scriptstyle\,\pm\, 12.36}$ \\
\bottomrule
\end{tabular}
\end{table}

\textbf{Impact of $E$.} Table \ref{tab:local_epochs} summarizes the results of varying $E$ on model accuracy under IID and non-IID settings. As shown in Table~\ref{tab:local_epochs}, under IID conditions, varying $E$ has only a minor effect on accuracy, with all configurations achieving consistently high performance around $99\%$. This observation is similar to the $C$ parameter. In contrast, under non-IID settings, the effect of $E$ is more pronounced and non-monotonic. When $\alpha$ = 0.1, increasing $E$ from 1 to 7 leads to a steady improvement in accuracy, with the best performance achieved at $E = 7$. However, further increasing $E$ results in a decline in accuracy. This behavior is even more evident when $\alpha$ = 0.01, where performance peaks at $E = 5$ and then drops sharply as $E$ increases, with $E = 11$ yielding the worst results. 

We attribute this behavior to the local models overfitting to their local data at higher values of $E$. Increasing $E$ allows clients to better optimize their local objectives. However, under non-IID data, larger values of $E$ cause local models to overfit to their skewed local distributions, leading to updates that deviate significantly from each other and the global objective. As a result, the aggregated model generalizes poorly. Hence, our results suggest that moderate values of $E$ should be preferred under non-IID settings.

\textbf{Impact of $K$.} We report the accuracy impact of varying $K$ in Table \ref{tab:num_clients}. In general, there is no strong relationship between $K$ and the model performance. Under IID conditions, the accuracies under varying $K$ are close to one another. Under non-IID conditions, the relationship becomes non-trivial and inconsistent. For both $\alpha$ = 0.1 and $\alpha$ = 0.01, accuracy peaks at $K$ = 25, while both smaller and larger $K$ values result in reduced accuracy. In particular, for larger $K$, both the occurrence likelihood and the probability of selecting highly skewed clients in each round increases, which negatively affects model accuracy. 

\section{Conclusion} \label{sec:Conclusion}

\begin{table}[!t]
\centering
\caption{Impact of the number of clients ($K$) on model accuracy}
\label{tab:num_clients}
\begin{tabular}{cccc}
\toprule
 \textbf{$K$} & \textbf{IID} & \textbf{$\alpha = 0.1$} & \textbf{$\alpha = 0.01$}  \\
\midrule
10  & $99.14 {\scriptstyle\,\pm\, 0.25}$ & $84.94 {\scriptstyle\,\pm\, 17.80}$ & $63.35 {\scriptstyle\,\pm\, 15.60}$ \\
25  & $\mathbf{99.26} {\scriptstyle\,\pm\, 0.11}$ & $\mathbf{88.84} {\scriptstyle\,\pm\, 4.72}$ & $\mathbf{69.05} {\scriptstyle\,\pm\, 15.55}$ \\
50  & $99.04 {\scriptstyle\,\pm\, 0.28}$ & $83.92 {\scriptstyle\,\pm\, 9.03}$  & $47.63 {\scriptstyle\,\pm\, 22.96}$ \\
100 & $98.93 {\scriptstyle\,\pm\, 0.19}$ & $82.91 {\scriptstyle\,\pm\, 13.79}$ & $58.70 {\scriptstyle\,\pm\, 3.68}$ \\
\bottomrule
\end{tabular}
\end{table}

In this paper, we studied FL for malware image classification under data heterogeneity (non-IID). Through an empirical study across three datasets (MalImg, Virus-MNIST, and MalNet), we showed that both accuracy and training stability degrade as data becomes increasingly non-IID. We introduced a set of round-level statistics to characterize the data contributed by participating clients per FL round, and observed that class entropy exhibits the strongest correlation with global model accuracy. Leveraging this insight, we proposed an entropy-aware client selection strategy that enforces high class entropy among participating clients, and showed that it outperforms standard random client selection.

The entropy-aware client selection mechanism proposed depends on server access to client data statistics. Future work could investigate privacy-preserving estimation of the statistics in section 4.4, such as inference from model updates or utilizing differential private mechanisms. Combining distribution-aware client selection with robust FL aggregation algorithms, and studying their impacts on malware image classification is another promising direction.

\section*{\uppercase{Acknowledgements}}

This research was supported by the Scientific and Technological Research Council of Türkiye (TÜBİTAK) under 2247-D Award 123C595, the BAGEP Outstanding Young Scientist Award, TÜBİTAK 2247-A Award 121C338, and EU CHIST-ERA TÜBİTAK Grant 125N724. The authors thank TÜBİTAK and the Science Academy for their support.


\bibliographystyle{apalike}
{\small
\bibliography{references}}

\end{document}